\documentclass{aa} 
\usepackage[varg]{txfonts}
\usepackage{graphicx}
\usepackage{url}
\usepackage{booktabs}
\usepackage[hyphenbreaks]{breakurl}
\usepackage[colorlinks=true,allcolors=blue]{hyperref}
\usepackage{orcidlink}

\def\Sg{$\Sigma_{\rm gas}$}
\def\S*{$\Sigma_{\rm SFR}$}
\def\Scii{$\Sigma_{\rm [CII]}$}

\def\alphacii{\alpha_{\rm [CII]}}
\def\alphaciir{W_{\rm [CII]}}
\def\unitswcii{(\rm M_{\odot} \, kpc^{-2})/( L_{\odot} kpc^{-2})}
\defcitealias{ferrara2019}{F19}
\begin{document} 
\title{Spatially resolved [CII]-gas conversion factor in early galaxies}
\author{
      L. Vallini\thanks{\email{livia.vallini@inaf.it }}\inst{1}\orcidlink{0000-0002-3258-3672}
      \and 
      A. Pallottini\inst{2,3}\orcidlink{0000-0002-7129-5761} 
      \and 
      M. Kohandel\inst{3}\orcidlink{0000-0003-1041-7865}
      \and 
      L. Sommovigo\inst{4}\orcidlink{0000-0002-2906-2200}
      \and
      A. Ferrara\inst{3}\orcidlink{0000-0002-9400-7312}
      \and 
      M. Bethermin\inst{5}\orcidlink{0000-0002-3915-2015}
      \and
      R. Herrera-Camus\inst{6}\orcidlink{0000-0002-2775-0595}
      \and
      S. Carniani\inst{3}\orcidlink{0000-0002-6719-380X}
      \and
      A. Faisst\inst{7}\orcidlink{0000-0002-9382-9832}
      \and
      A. Zanella\inst{1}\orcidlink{0000-0002-3915-2015}
      \and
      F. Pozzi\inst{8,1}\orcidlink{0000-0002-7412-647X}
      \and
      M. Dessauges-Zavadsky\inst{9}\orcidlink{0000-0003-0348-2917}
      \and
      C. Gruppioni\inst{1}\orcidlink{0000-0002-5836-4056}
      \and
      E. Veraldi\inst{10}\orcidlink{0009-0007-1304-7771}
      \and
      C. Accard\inst{5}\orcidlink{0009-0005-9982-7239}    } 
\institute{
      INAF, Osservatorio di Astrofisica e Scienza dello Spazio, Via P. Gobetti 93/3, I-40129, Bologna, Italy
      \and  Universit\'{a} di Pisa, Dipartimento di Fisica ``Enrico Fermi'', Largo Bruno Pontecorvo 3, Pisa I-56127, Italy
      \and Scuola Normale Superiore, Piazza dei Cavalieri 7, I-56126 Pisa, Italy
      \and Center for Computational Astrophysics, Flatiron Institute, 162 Fifth Avenue, New York, NY 10010, USA
      \and Universit\'e de Strasbourg, CNRS, Observatoire astronomique de Strasbourg, UMR 7550, 67000 Strasbourg, France
      \and Departamento de Astronom\'ia, Universidad de Concepci\'on, Barrio Universitario, Concepci\'on, Chile
      \and Caltech/IPAC, 1200 E. California Blvd. Pasadena, CA 91125, USA 
      \and University of Bologna, Department of Physics and Astronomy “Augusto Righi”, Via Gobetti 93/2, I-40129 Bologna, Italy
      \and D\'epartement d'Astronomie, Universit\'e de Gen\`eve, Chemin Pegasi 51, 1290 Versoix, Switzerland
      \and Scuola Internazionale Superiore Studi Avanzati (SISSA), Physics Area, Via Bonomea 265, I-34136 Trieste, Italy
      }

\titlerunning{Spatially resolved [CII]-gas conversion factor}
\authorrunning{Vallini et al.}

\date{}


\abstract
{} 
{Determining how efficiently gas collapses into stars at high-redshift is key to understanding galaxy evolution in the Epoch of Reionization (EoR). Globally, this process is quantified by the gas depletion time ($t_{\rm dep}$); on resolved scales, by the slope and normalization of the Kennicutt-Schmidt (KS) relation. This work explores the global ($\alpha_{\rm [CII]}$) and spatially resolved ($W_{\rm [CII]}$) [CII]-to-gas conversion factors at high-$z$ and their role in inferring reliable gas masses, surface densities, and $t_{\rm dep}$ in the EoR.}
{We select galaxies at $4<z<9$ from the \texttt{SERRA} cosmological zoom-in simulation, that features on-the-fly radiative transfer and resolves interstellar medium properties down to $\approx 30\,\rm pc$. The [CII] emission modelling from photodissociation regions allow us to derive global $\alphacii$, and maps of $\alphaciir$. We study their dependence on gas metallicity ($Z$), density ($n$), Mach number ($\mathcal{M}$), and burstiness parameter ($\kappa_s$), and provide best fit relations.}
{The $\alpha_{\rm [CII]}$ decreases with increasing $Z$ and galaxy compactness, while the resolved $W_{\rm [CII]}$ shows two regimes: at $Z< 0.2 Z_\odot$, it anticorrelates with n and Z, but not with $\kappa_s$; above this threshold, it also depends on $\kappa_s$, with more bursty regions showing lower conversion factors. This implies $W_{\rm [CII]}\propto \Sigma_{\rm [CII]}^{-0.5}$, as dense, metal-rich, and bursty regions exhibit higher [CII] surface brightness. Applying a constant $\alpha_{\rm [CII]}$ overestimates $\Sigma_{\rm gas}$ in bright $\Sigma_{\rm [CII]}$ patches, thus flattening the KS slope and overestimating $t_{\rm dep}$ by up to 4$\times$.}
{}

\maketitle
%
\section{Introduction}
The James Webb Space Telescope (JWST) has revealed an unexpected abundance of ultraviolet (UV)-luminous, massive galaxies in the Epoch of Reionization \citep[EoR; e.g.,][]{naidu2022, finkelstein2022, castellano2023, arrabal2023, tacchella2023, adams2024, donnan2024, finkelstein2024, carniani2024}, chal\-len\-ging our understanding of early star formation and galaxy evolution.
The large stellar masses, number density, and blue colors of newly discovered sources do not easily fit within pre-JWST frameworks, hence, new models have been proposed to explain these results. Leaving aside modifications of $\Lambda$CDM \citep[e.g.][]{padmanabhan2023, matteri2025, matteri2025_b}, the list of attempts includes enhanced star formation efficiency due to delayed feedback \citep{dekel2023, li2024}, top-heavy initial mass function \citep[e.g.][]{hutter2024, trinca2024}, dust removal driven by radiation-driven outflows which reduce the impact of dust on UV attenuation \citep{ferrara2023, ferrara2024}, and stochastic variations in galaxy luminosity caused by bursty star formation \citep[e.g.][]{pallottini2023, sun2023, mirocha2023, nikolic2024}. 

Most of these scenarios are linked to how quickly gas is converted into stars, as quantified by the depletion time ($t_{\rm dep} = M_{\rm gas} / \rm SFR$), thus deriving the cold gas mass ($M_{\rm gas}$) that fuels star formation in high-$z$ galaxies, and their star formation rate (SFR), is of utmost importance.
A powerful tracer of the cold gas that eventually collapses to form stars is the fine-structure transition of ionized carbon (C$^+$) at 158 $\mu$m ([CII]). This is the brightest line in the rest-frame far-infrared (FIR) and, albeit up to $\approx 10\%-20\%$ of the flux can originate diffuse ionized gas \citep{croxal2017}, it predominantly traces the diffuse cold neutral medium \citep[CNM, $T\approx50-100$ K;][]{wolfire2003} and dense photodissociation regions \citep[PDRs;][]{wolfire2022} representing the external layers of giant molecular clouds (GMCs). Since the advent of the Atacama Large Millimeter Array (ALMA), the [CII] 158 $\mu$m line has been detected both in large galaxy samples (\citealt{bethermin2020, faisst2020} for ALPINE, \citealt{bouwens2022} for REBELS, and \citealt{herrera-camus2025} for CRISTAL surveys, respectively) and in targeted sources \citep[e.g.,][]{maiolino2015, capak2015, knudsen2016, carniani2018, matthee2019, glazer2024} in the EoR.\\

Interestingly, JWST-selected sources, with spectroscopic redshifts $z>10$, remain elusive in [CII] observations, with only upper limits obtained so far \citep{fudamoto2024, schouws2025}.
While too short of observing times can be an issue \citep{kohandel2024}, a plausible possibility is the rapid exhaustion of cold gas in their ISM thanks to vigorous burst of star formation \citep[][]{witstok2025}. Precisely determining $t_{\rm dep}$ is thus critical, but it relies on the adopted [CII]-to-$M_{\rm gas}$ conversion factor ($\alpha_{\text{[CII]}}\equiv M_{\rm gas}/L_{\rm [CII]}$), which introduces systematic uncertainties in $M_{\rm gas}$\footnote{In this paper we adopt $\alphacii$ as \textit{total cold gas} conversion factor without distinguishing between molecular and atomic. As detailed in Sec. \ref{sec:cii_emission}, the [CII] emission is indeed computed accounting for both the outer PDR layer and the fully molecular clumps within molecular clouds. } estimates \citep{sommovigo2021, mitsuhashi2025, algera2025}. \citet{zanella2018} first proposed [CII] as an alternative to the less luminous CO to pin-down \textit{molecular} gas in high-$z$ sources \citep[see][for low-$z$]{wolfire2010, pineda2014}, finding $\alphacii = (31 \rm \, M_{\odot}\, L_{\odot}^{-1})$. Other studies reported a wide range of values, ranging from $\alpha_{\text{[CII]}} \approx 72 \, \rm \, M_{\odot}\, L_{\odot}^{-1}$ for local metal poor dwarf galaxies (\citealt{madden2020}, but see also \citealt{ramambason2024}), to significantly lower values ($\alpha_{\text{[CII]}} \approx 5 \, - 10\, \rm M_{\odot} L_{\odot}^{-1}$) in gravitationally lensed dusty star-forming galaxies at $z \sim 4.5$ \citep{rizzo2021}, and quasars at $z \approx 6-7$ \citep{kaasinen2024, salvestrini2025}. At $z\approx 4-5.5$, \citet{dessauges-zavadsky2020} found agreement between the dynamical mass of galaxies in ALPINE and that inferred from [CII] adopting the \citet{zanella2018} conversion factor, thus supporting the [CII] as a total cold gas mass tracer \citep[see also][focussing on HI]{heintz2021}.

We are living in an era when JWST and ALMA can spatially resolve EoR galaxies in both the rest-frame optical/UV \citep[e.g.,][]{ubler2023, tripodi2024, venturi2024, jones2024} and the FIR \citep[e.g.,][]{witstok2022, posses2023, vallini2024, ikeda2024, solimano2024}. As a result, it is now possible to measure the resolved Kennicutt–Schmidt (KS) relation \citep[$\Sigma_{\rm SFR } \propto \kappa_s \Sigma_{\rm gas}^{1.4}$][]{schmidt1959, kennicutt1998} and thereby constrain the local gas depletion time ($t_{\rm dep}= \Sigma_{\rm gas}/\Sigma_{\rm SFR } \propto 1/\kappa_s$).The depletion time is inversely proportional to the so-called "burstiness parameter" \citep[first introduced by][]{ferrara2019} namely the offset ($\kappa_s$) from the KS relation.  
Crucially, it remains unclear whether using a single $\alpha_{\text{[CII]}}$ for different sub-kpc regions in the ISM of high-$z$ sources provides an accurate determination of the gas surface density ($\Sigma_{\rm gas}$) from the [CII] 158 $\mu$m surface brightness ($\Sigma_{\rm [CII]}$) \citep[e.g.,][Accard in prep.]{bethermin2023, vallini2024}. Sub-kpc variations in metallicity, density, and turbulence can significantly impact [CII] emission \citep[e.g.,][]{vallini2015, ferrara2019, pallottini:2019, herrera-camus2021, veraldi2025}, thus a spatially resolved conversion factor, $\alphaciir \equiv \Sigma_{\rm gas}/\Sigma_{\rm [CII]}$, is likely required.\\

This work aims to theoretically shed light on the connection between $\alphaciir$ and the underlying ISM conditions in early galaxies, and on the relation between $\alphaciir$ and $\alphacii$. Theoretical works focusing on the [CII] emission both from resolved GMCs in the local Universe \citep{ebagezio2023, gurman2024} and from galaxies in the EoR \citep{vallini2015, olsen2017,lagache2018, vizgan2022, casavecchia2025}, have gained significant traction over the past decade. Among the many approaches, cosmological zoom-in simulations featuring on-the-fly radiative transfer and resolving $\approx 10\rm \, pc$ scales, to accurately model [CII] emission from diffuse neutral gas, PDRs, and GMCs \citep[e.g.,][]{katz2017, katz2022, lupi2020, pallottini:2019, pallottini2022,schimek2024} are the most advanced tools for studying the ISM properties of high-redshift galaxies, while retaining the cosmological context. In this paper, we exploit the state-of-the-art \texttt{SERRA} cosmological zoom-in simulations \citep{pallottini2022}. 

The paper is structured as follows: in Section \ref{sec:serra} we briefly describe \texttt{SERRA}, the [CII] emission model, and the sample of galaxies considered in this work. In Sections \ref{sec:globalalpha} and \ref{sec:spatiallyalpha} we present our results regarding $\alphacii$, $\alphaciir$ and their connection with the ISM properties. In Section \ref{sec:discussion} we discuss the observational implications for the derivation of the total cold gas mass, the $t_{\rm dep}$, and the KS relation. The conclusions are outlined in Section \ref{sec:conclusions}.
\section{SERRA simulation}
\begin{figure*}
    \centering
    \includegraphics[width=0.41\textwidth]{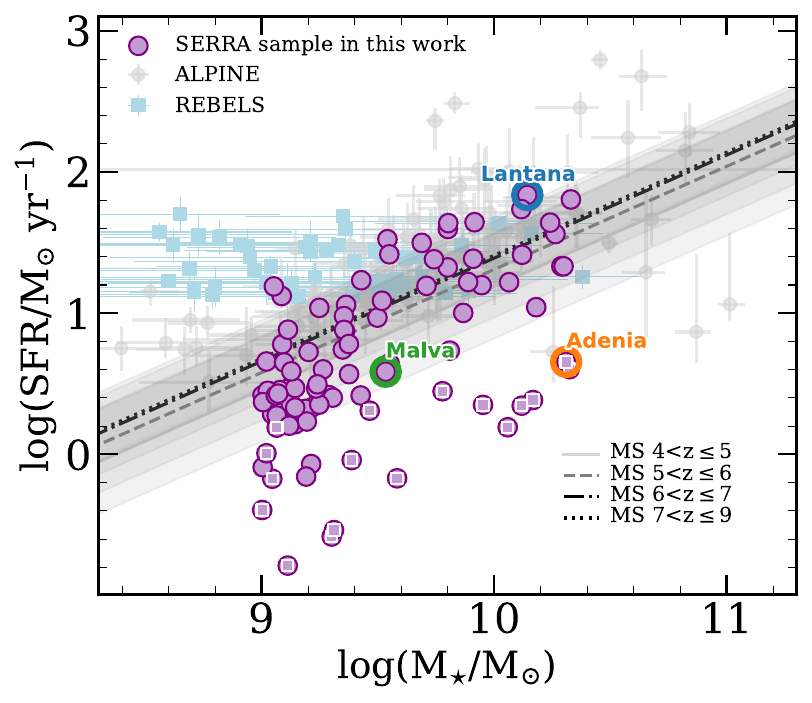}
    \includegraphics[width=0.42\textwidth]{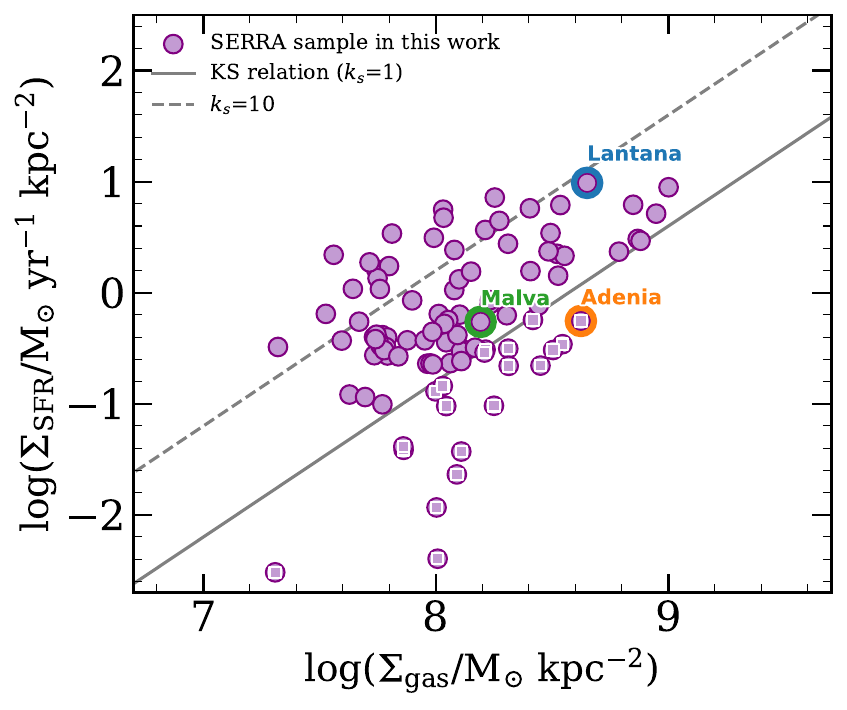}
    \caption{Left panel: SFR vs M$_{\star}$ relation for the \texttt{SERRA} sample (purple points), along with the main sequence parametrization in the CEERS sample in different redshift bins (gray lines), with their $1\sigma$ scatter (grey shaded areas), for 100 Myr-averaged SFR \citep{cole2025}. The SFR vs M$_{\star}$ of ALPINE \citep[][grey points]{khusanova2021}, and REBELS \citep[][light blue squares]{inami2022, topping2022} are also plotted for comparison. Right panel: $\Sigma_{\rm SFR}$ vs $\Sigma_{\rm gas}$ KS relation for the \texttt{SERRA} galaxies. The local KS relation is plotted with a solid gray line, while that obtained for burstiness parameter $\kappa_s=10$ is shown with a dashed gray line. Those sources that were below the KS relation (points with white inner square) fall also below the MS. In both plots we highlight the three example galaxies discussed in Section \ref{sec:globalalpha} and whose cutouts are presented in Figure \ref{fig:maps}.
    \label{fig:sample}
    }
\end{figure*}
\label{sec:serra}
\subsection{Galaxy formation and evolution}
The \texttt{SERRA} simulation suite investigates the processes driving galaxies' formation and evolution during the EoR \citep{pallottini2022}. \texttt{SERRA} uses a tailored version of the adaptive mesh refinement (AMR) code \textsc{Ramses} \citep{teyssier:2002} to evolve gas and dark matter (DM). The simulations start from $z=100$, where initial conditions are generated
with MUSIC \citep{hahn2011}, in a cosmological volume of (20 Mpc/h)$^3$. \textsc{Ramses} is used to track the evolution of DM, stars, and gas, reaching a
baryon mass resolution of $1.2 \times 10^4 \, M_{\odot}$ and spatial resolution of $\approx 30$ pc at $z\approx 6$ in the zoom-in regions, i.e., about the mass and size of typical GMCs \citep{federrath2013}.

\begin{figure*}
    \centering
    \includegraphics[width=0.3\linewidth]{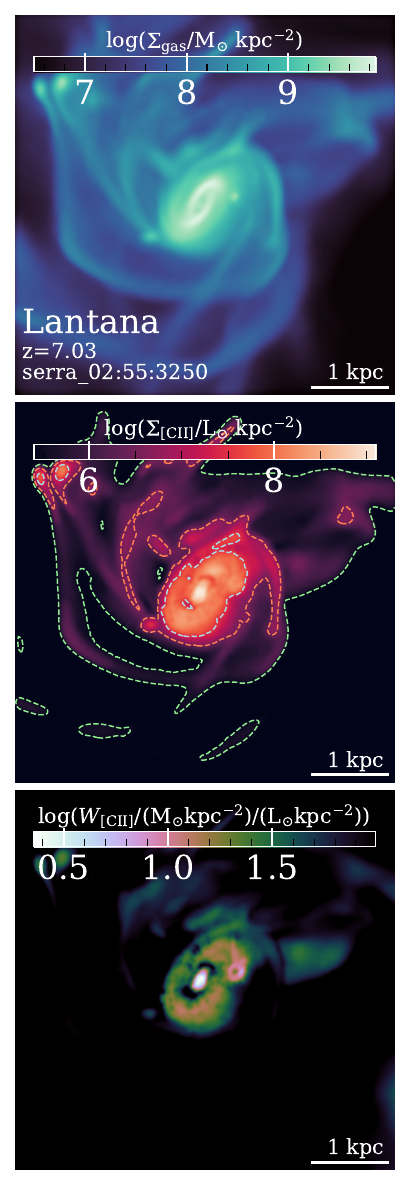}
    \includegraphics[width=0.3\linewidth]{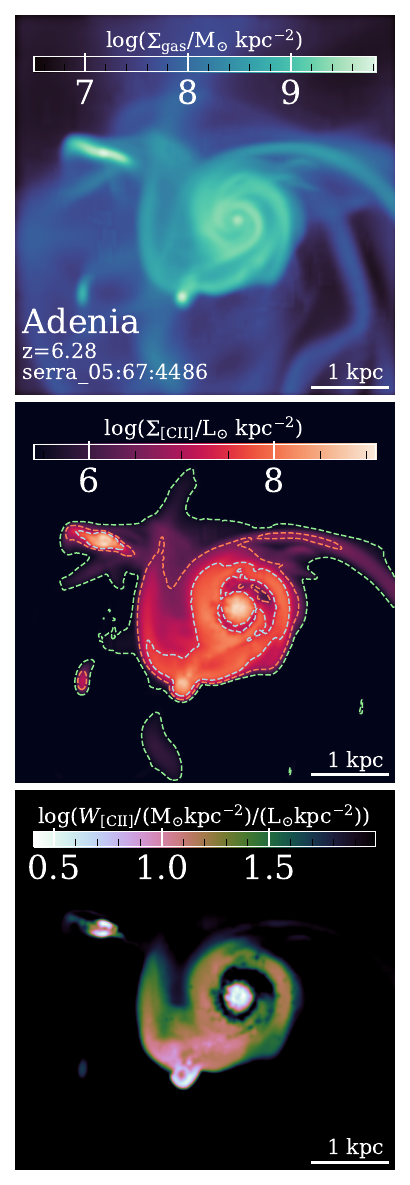}
    \includegraphics[width=0.3\linewidth]{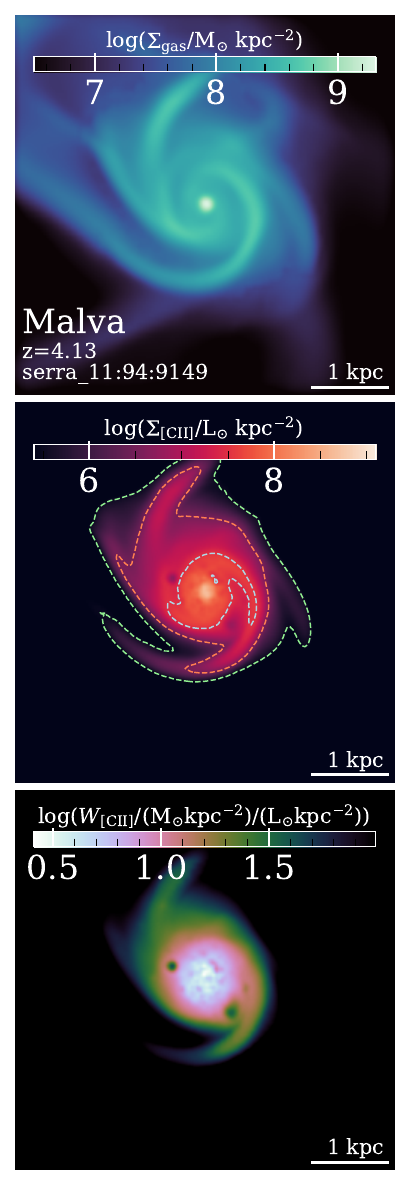}
    \caption{Cutouts of $5\times 5$ kpc$^2$ size centered on three \texttt{SERRA} galaxies. From left to right: Lantana (\texttt{serra02:55:3350}, $z\approx 7.0$), Adenia (\texttt{serra05:67:4486}, $z\approx 6.3$), and Malva (\texttt{serra11:94:9149}, $z\approx 4.1$). From top to bottom: gas surface density, [CII] surface brightness, and spatially resolved conversion factor, $W_{\rm [CII]}$, in units of [$\unitswcii$]. To guide the eye, dashed green, orange, and light blue lines in the \Scii~maps highlight $\Sigma_{\rm [CII]}=10^{5.5}, 10^{6.5}, 10^{7.5}\, \rm L_{\odot}\, kpc^{-2}$,  respectively. }
    \label{fig:maps}
\end{figure*}

The chemical evolution is handled with \textsc{Krome} \citep{grassi:2014mnras}, which solves a non-equilibrium chemical network encompassing species such as H, H\textsuperscript{+}, H\textsuperscript{-}, He, He\textsuperscript{+}, He\textsuperscript{++}, H\textsubscript{2}, H\textsubscript{2}\textsuperscript{+}, electrons, and metals, with roughly 40 reactions \citep{bovino:2016aa, pallottini2017}. Metallicity ($Z$) is tracked by summing heavy elements under the assumption of solar abundance ratios for metal species \citep{asplund2009}. Dust scales with $Z$ as, $\mathcal{D}=Z  \mathcal{D_{\odot}}$, where $\mathcal D_{\odot}=0.3$ is the dust to metal ratio at solar metallicity \citep{remiruyer2014}. \texttt{SERRA} also adopts a Milky Way-like dust composition and
grain size distribution \citep{weingartner2001}.
Star formation follows a KS-like relation \citep{schmidt1959, kennicutt1998}, with H$_2$ being the primary fuel for star formation \citep{pallottini2017}. 
The H$_2$ density is obtained from the non-equilibrium chemical network, and taking into account of the photo-dissociation produced by the Lyman Werner flux, computed on-the-fly \citep{pallottini:2019}. \citet{lupi2020b} showed that, adopting the same initial conditions but a SF recipe based on the local gas turbulence, results in differences in the stellar mass build-up only during the first $\approx100$ Myr of evolution. In this case the SFR history is more prone to bursts, depending also on the stellar feedback adopted \citep[see also][]{pallottini:2025}.
Stars inject metals, radiation, and mechanical energy \citep{pallottini2017b}. 
\texttt{SERRA} treats feedback both in thermal and turbulent forms: thermal energy cools via \textsc{Krome}-modeled thermo-chemical evolution, while turbulent energy dissipates over an eddy turnover timescale \citep{macLow1999}. Finally, the interstellar radiation field (ISRF) is coupled with the chemical processes \citep{pallottini:2019} and evolves dynamically with 5 frequency bins, as it is solved on-the-fly using the moment-based RT module in \textsc{Ramses-rt} \citep{rosdahl2013}. The first two
bins cover the \citep{habing1968} band (6.0-13.6eV), with the latter bin being specific for the Lyman–Werner radiation (11.2-13.6 eV), which can photodissociate H$_2$. The remaining three bins cover the H-ionizing photons up to the first ionization level of He (13.6–24.59 eV). Further details can be found in \citet{pallottini:2019}.

\subsection{[CII] emission}
\label{sec:cii_emission}

The [CII] emission\footnote{Several other emission lines, besides [CII], are computed in \texttt{SERRA}, both in the rest-frame FIR and in the optical.} is computed on a cell-by-cell basis using grids of \textsc{Cloudy} models \citep{ferland2017}. 
In \texttt{SERRA}, both species by species abundances and line emission are post-processed to improve the precision of the prediction, at the cost of losing track of non-equilibrium effects. The opposite approach -- adopting the chemistry from the simulation for line predictions at the cost of precision -- is possible, and shows very little difference for the [CII]  \citep[see][]{lupi2020b}.

The \texttt{Cloudy} models used for the post processing account for variations in density, metallicity, and radiation field intensity, with the column density ($N_H$) serving as the stopping criterion for the calculations. The PDR ($N_H \approx 10^{21.5} \rm cm^{-2}$) and mo\-le\-cu\-lar layers ($N_H \approx 10^{22.5}\rm cm^{-2}$) are fully sampled in the calculations. The SED used for the radiation field incident on the gas slab in \textsc{Cloudy} assume a 10 Myr old \textsc{Starburst99} \citep{leitherer1999} stellar population. We have two identical grids of \texttt{Cloudy} models, in which the impinging SED includes -- does not include -- ionizing radiation ($h\nu>13.6$ eV). Each grid, equally spaced in log scale, comprises 17 density bins ($10^{-2} \leq n/{\rm cm}^{-3} \leq 10^{6.5}$), 8 metallicity bins ($10^{-3} \leq Z/Z \leq 10^{0.5}$), and 12 ISRF intensity bins ($10^{-1} \leq G/G_0 \leq 10^{4.5}$, with $G_0=1.6 \times  10^{-3}\rm erg\, s^{-1}\, cm^{-2}$ in the $6 \leq  h\nu < 13.6\rm \, eV$ \citealt{habing1968} band), resulting in 1632 models.
Due to the $\sim$30~pc resolution of \textsc{SERRA}, dense H\,\textsc{ii} regions, GMCs, and PDRs remain unresolved and require a subgrid treatment \citep[see][]{decataldo2017, vallini2018, pallottini:2019}. [CII] emission is computed using the \texttt{Cloudy} grids outlined above, interpolated over $G_0$, $N_\mathrm{H}$, and $Z$ in the cell. We adopt \texttt{Cloudy} models with ionizing radiation either if the HII region is resolved on-the-fly in the simulation or if young ($t_\star < 10$ Myr) stars are contained the gas cell. If such conditions are not met, the radiation is treated as non-ionizing. For the interpolation in gas density, we account for the log-normal density distribution within each cell, following the approach described in \citet{vallini2018} where the sub-grid density PDF is parametrized as a function of the local Mach number ($\mathcal{M}$), defined as the ratio of the turbulent plus thermal velocities over the turbulent velocity.
The results of the \textsc{Cloudy} postprocessing are spatially resolved maps and HyperSpectral Data Cubes (HSDCs), which provide line spectra in position–position–velocity space \citep[][]{Kohandel2020, kohandel2024}. 
For optically thin lines, such as the [CII] 158 $\mu$m, the luminosity is the sum of the emission cells within the HSDC centered on each galaxy and with side of 5 kpc.

\subsection{Sample selection}

We start from a parent sample of 3218 \texttt{SERRA} galaxies ($10^8 \rm M_{\odot}<M_{\star}<10^{10.3} M_{\odot}$), whose kinematic properties in both [CII] and H$\alpha$ were analyzed in \citet{kohandel2024}. For this work, we select a subsample of 98 sources with stellar masses $M_\star>10^9 \rm M_{\odot}$, in the redshift range $4.0\leq z \leq 8.9$. Our sample includes galaxies with SFR$=0.01-70\,\rm M_{\odot}\, yr^{-1}$. The SFR is averaged over the last 30 Myr.
In Figure \ref{fig:sample}, we show the location of our sources in the SFR–$M_\star$ and $\Sigma_{\rm gas}-\Sigma_{\rm SFR}$ planes. In the former, the galaxies broadly follow the 'star formation main sequence' (MS), although defining the MS is not straightforward for sources with bursty star formation \citep{gelli2025}, particularly considering its evolution across a broad range of redshifts. In Figure \ref{fig:sample}, we report the \citet{cole2025} parametrization of the MS in different redshift bins encompassing redshift range of our sample. From Fig. \ref{fig:sample}, it is clear that overall, \texttt{SERRA} sources are scattered along and below the \citet{cole2025} MS parametrizations. For comparison we also report the location of the two largest [CII] detected samples at high-$z$: ALPINE \citep[see][]{khusanova2021}, and REBELS \citep[see][]{inami2022,topping2022}.

In the right panel of Fig. \ref{fig:sample} the major fraction of our galaxies lie on and above the local KS relation \citep{heiderman10, delosreyes2019} by a factor of up to $\kappa_s \approx 10$ \citep[see also][]{pallottini2022}. Finally, we note that the sources below the MS are also below the local KS relation. These galaxies are experiencing quenching in star formation over the last 30 Myr, yet their gas is still not completely exhausted ($\Sigma_{\rm gas}\gtrsim 10^8\,\rm M_{\odot} kpc^{-2}$). How this translates into the [CII]-to-$M_{\rm gas}$ conversion factor is addressed in the next sections.

\section{The global $\alphacii$ conversion factor in SERRA}
\label{sec:globalalpha}

The $\alphacii=M_{\rm gas}/L_{\rm [CII]}$ (in units of $M_{\odot} L_{\odot}^{-1}$) is computed from the $M_{\rm gas}$ and $L_{\rm [CII]}$ obtained by integrating the gas surface density, \Sg, and the [CII] surface brightness, $\Sigma_{\rm [CII]}$, within a $5\times5\rm\, kpc^2$ field of view (FOV) centered on each simulated galaxy which is obtained collapsing (i.e. projecting in 2D) the HSDCs.
\begin{figure}
    \centering
   \includegraphics[width=0.46\textwidth]{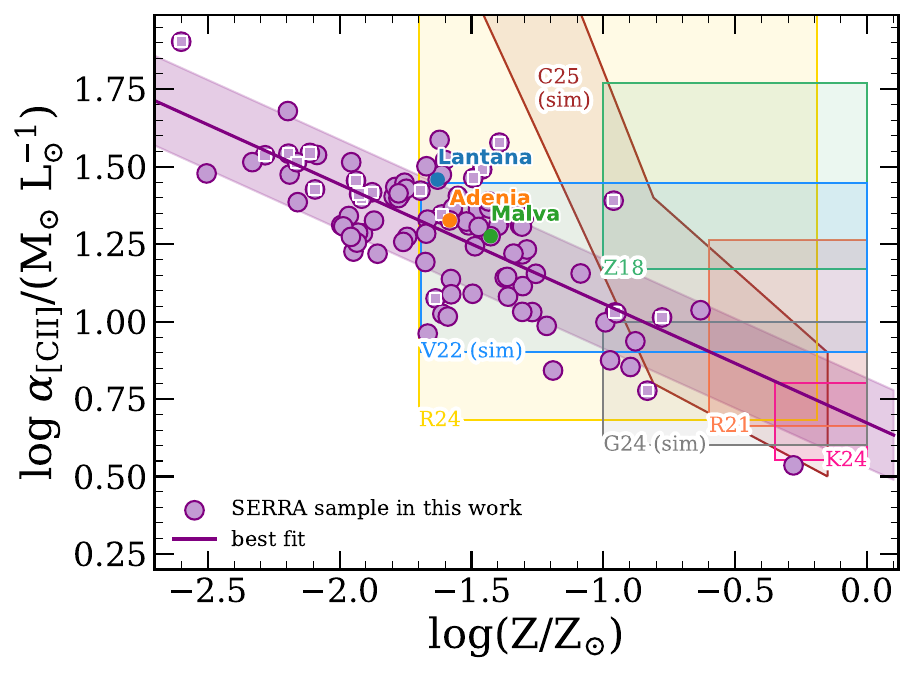}
   \caption{The $\alphacii$ vs $\log Z$ for the \texttt{SERRA} galaxies (purple points, with inner white square if they fall below the local KS relation) along with the best fit linear regression $\pm 1\sigma$ dispersion (purple line and shaded area). Lantana, Adenia, and Malva are plotted in blue, orange, and green respectively. Previous derivations of $\alphacii$ in literature are shown with shaded areas as follows: \citet{zanella2018} green shaded area ($\log \alphacii =1.5$ with 0.3 dex dispersion over $\log (Z/Z_{\odot})\approx [-1, 0.0]$); \citet{ramambason2024} yellow shaded area (converted into $M_\odot L_\odot^{-1}$ units, over $\log (Z/Z_{\odot})\approx [-1.69, -0.3]$ range); \citet{rizzo2021} ($\log \alphacii =[0.66, 1.3]$ assuming a fiducial range $\log (Z/Z_{\odot})\approx [-0.5, 0.0]$ for the DSFG galaxies in the sample), orange shaded area; \citet{kaasinen2024} that report $\alphacii = 4.6 \pm 1.6 \times$ lower than \citet{zanella2018}, over $\log (Z/Z_{\odot})\approx [-0.3, 0.0]$ for the QSOs in their sample, pink shaded box. Cosmological and single cloud simulations (labelled "sim") from \citet{vizgan2022, gurman2024} are almost constant within their $Z$ range, and are plotted in blue and grey, respectively. The $\alphacii$-$Z_\star$ relation by \citet{casavecchia2025} is shown in brown.}
    \label{fig:alphaciidistrib}
\end{figure}
In Figure \ref{fig:maps} we show example cutouts of \Sg, and \Scii~for three sources spanning redshifts from well within the EoR  (Lantana, \texttt{serra02:55:3350}, at $z=7.03$) to the end of reionization (Adenia, \texttt{serra05:67:4486}, at $z=6.28$), and half a Gyr after the EoR (Malva, \texttt{serra11:94:9149} at $z=4.13$). 

The mean conversion factor in the \texttt{SERRA} sample is $\log(\alphacii/M_{\odot} \,L_{\odot}^{-1})$ = 1.28\,  ($\sigma = \pm 0.2$ dex) and falls within the range bracketed by previous estimates in the literature.
We first explore correlations between $\alphacii$-redshift, $\alphacii$-$M_{\star}$, and $\alphacii$-$\Delta_{MS}$ (deviation from the MS). Our aim is to test whether the observational results from \citet{zanella2018}, which indicate $\alphacii$ does not significantly depend on any of these quantities up to $z\approx 2$, also hold for the \texttt{SERRA} galaxies up to $z\approx 9$. To quantify the presence (or lack thereof) of correlations, we perform a Spearman's test finding rank correlation coefficients $r_s=0.3, 0.07, 0.09$, and corresponding $p$-values are $0.002, 0.47, 0.37$, for redshift, $\log M_{\star}$, and $\log \Delta_{MS}$\footnote{We adopt the MS by \citet{cole2025} in the corresponding redshift bin of each SERRA galaxy.}, respectively. Overall, we recover a very weak, marginally significant correlation with $z$, while the $\alphacii$ vs. $M_{\star}$, and $\alphacii$ vs. $\Delta_{MS}$ do not show any statistically significant correlation, in agreement with \citet{zanella2018} \citep[see also][]{accurso2017, zhao2024}. The interested reader can find the plots in Figure \ref{fig:nocorrelation} in Appendix \ref{sec:appendix}.

\begin{figure*}
    \centering
    \includegraphics[width=0.9\linewidth]{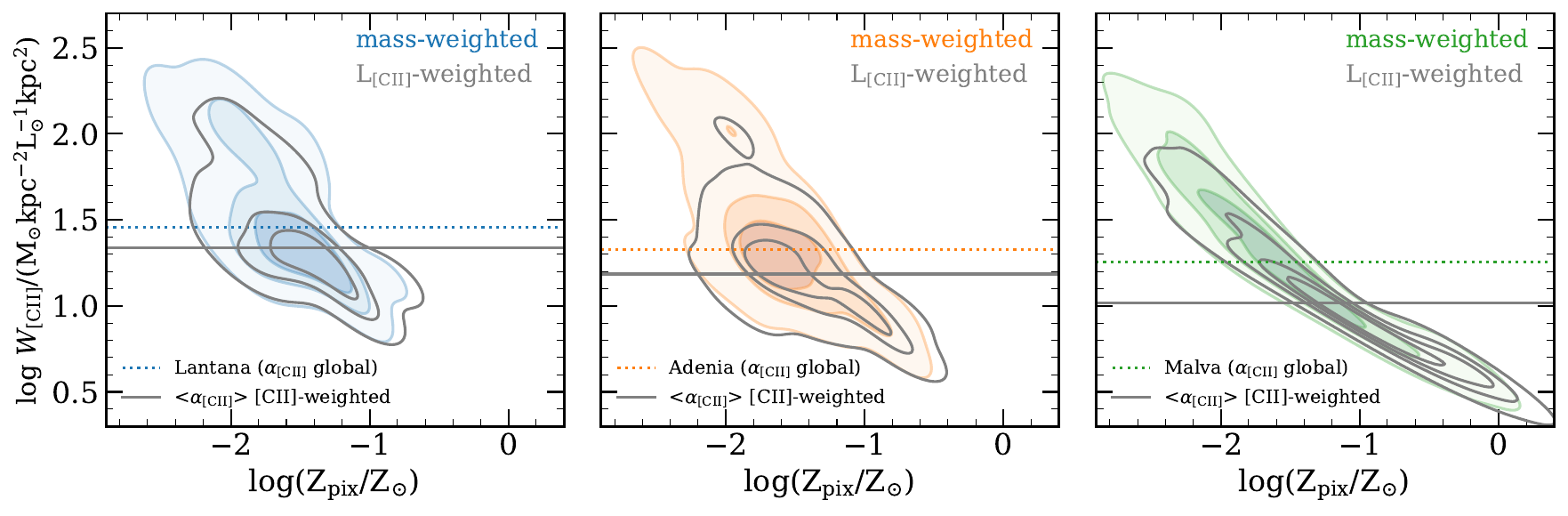}
    \includegraphics[width=0.9\linewidth]{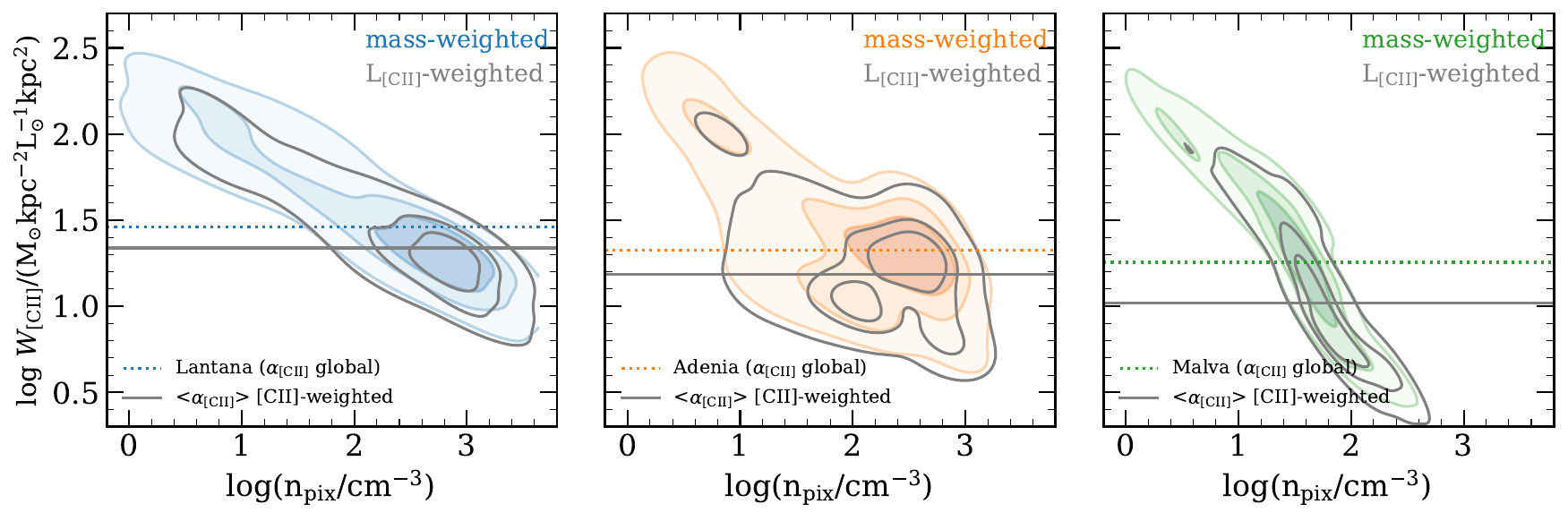}
    \includegraphics[width=0.9\linewidth]{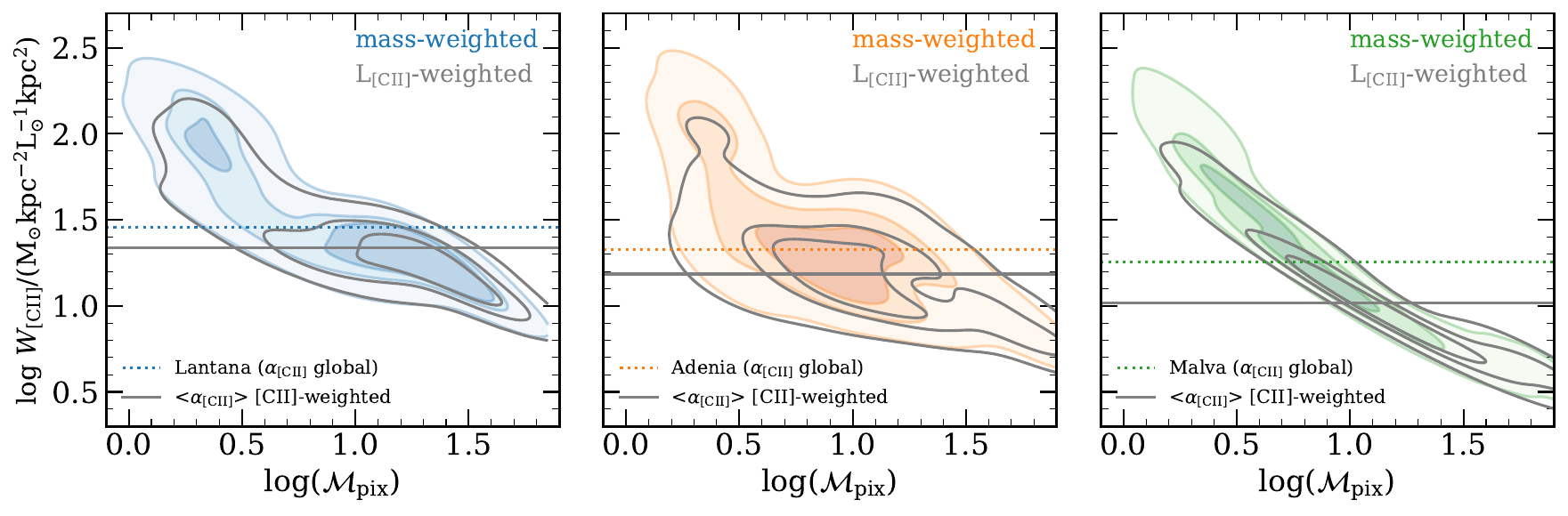}
    \caption{Spatially resolved conversion factor ($\alphaciir$) as a function of different physical properties within 30 pc size pixels in the ISM of the three galaxies shown in Fig. \ref{fig:maps}. From top to bottom: the mass-weighted KDE of pixels in the $\alphaciir$-$Z$ plane (top row), $\alphaciir$-$n$ plane (middle row), and $\alphaciir$-$\mathcal{M}$ plane (bottom row) for Lantana ($z=7.03$, left), Adenia ($z=6.28$, center), and Malva ($z=4.13$, right). In each panel, we also report the [CII]-weighted KDEs with grey contours. The global ([CII]-weighted) value of $\alphacii$ for each galaxy is reported with a dashed (solid) line.}
    \label{fig:2d_pdfs}
\end{figure*}

The situation is markedly different in the $\alphacii$–$Z$ plane (Figure \ref{fig:alphaciidistrib}), where we find a significant negative correlation ($r_s = -0.65$, $p$-value $= 5.1 \times 10^{-13}$) between $\alphacii$ and $Z$. We compare the \texttt{SERRA} trend with previous observational results spanning different metallicity ranges, as indicated by the shaded areas in Figure \ref{fig:alphaciidistrib}. In low-metallicity dwarf galaxies ($\log (Z/Z_{\odot}) \approx [-1.69, -0.3]$), \citet{ramambason2024} reported a median $\alpha_{\text{[CII]}} \approx 48 \,\rm M_\odot / L_\odot$, while the canonical value from \citet{zanella2018}, $\alphacii = 31 \,\rm M_{\odot}/L_{\odot}$, corresponds to sources with $\log (Z/Z_{\odot}) \approx [-1, 0]$. This appears somewhat in tension with our findings, as we measure $\alphacii \approx 5$–$10$ in the $0.5$–$1 \, Z_{\odot}$ range. \texttt{SERRA} values are instead consistent with the $\alphacii \approx 4$–$7.5 \,\rm M_{\odot}/L_{\odot}$ reported by \citet{rizzo2021} and \citet{kaasinen2024} for dusty star-forming and quasar host galaxies. While the metallicity of those sources is not well constrained, we adopt a fiducial $\log (Z/Z_{\odot}) \approx [-0.3, 0.0]$. 

In Figure \ref{fig:alphaciidistrib}, we also compare the \texttt{SERRA} results with both large-scale cosmological simulations \citep{vizgan2022, casavecchia2025} and GMC-scale simulations that incorporate non-equilibrium chemistry \citep{gurman2024}. \citet{vizgan2022} post-processed SIMBA simulations with SIGAME \citep{olsen2017}, finding $\alpha_{\text{[CII]}} \approx 18\, \rm M_\odot / L_\odot$ in galaxies at $z \approx 6$, which is in line with our results, albeit for galaxies with lower stellar masses. Note that they reported a very weak correlation with $Z$, in contrast to the significant negative correlation we find. \citet{casavecchia2025}, using the ColdSIM simulations, obtained a redshift-dependent molecular gas mass–to–[CII] luminosity relation. When extrapolated to match the median redshift and [CII] luminosity of the \texttt{SERRA} sample, their relation suggests $\alphacii \approx 28\, \rm M_\odot / L_\odot$. \citet{casavecchia2025} also report a negative correlation between $\alphacii$ and stellar metallicity in their simulation, as highlighted by the shaded C25 area in Figure \ref{fig:alphaciidistrib}. The $\alphacii$ values in \texttt{SERRA} are broadly consistent with those from GMC-scale simulations by \citet{gurman2024}, which span $Z \approx 0.1$–$1 \, \rm Z_{\odot}$. This supports the idea that radiative transfer and ISM properties on sub-pc scales must be considered, as they significantly impact the accuracy of [CII] emission predictions. Finally, an important caveat to be emphasized is that metallicity in \texttt{SERRA} is derived directly from the O/H ratio in each cell, then converted to solar units using \citet{asplund2009}. In contrast, observational estiamates of $Z$ rely on calibrations based on emission line ratios \citep[see][]{pallottini:2025}. A dedicated study comparing $Z$ derived from these calibrations with the intrinsic O/H values in \texttt{SERRA} is deferred to future work. 

The linear regression fit to our simulated sources is given by:
\begin{equation} 
\label{eq:alpha_Z_fit} 
\log (\alphacii/ M_{\odot} L_{\odot}^{-1}) = -0.39 \log (Z/Z_{\odot}) + 0.67 
\end{equation} 
with a dispersion of 0.14 dex. The errors on the best fit parameters are summarized in Table \ref{tab:best_fit_relations}. Interestingly, we note that the anticorrelation of $\alphacii$ with metallicity mirrors the well-known anticorrelation of the CO-to-H$_2$ conversion factor with $Z$ \citep[$\alpha_{\rm CO}\propto Z^{-0.5}$;][]{bolatto2013, narayanan2011}. At low $Z$, the reduced dust shielding allows UV radiation to penetrate deeper into GMCs, leading to CO photodissociation and to a drop in the CO luminosity efficiency, thus boosting $\alpha_{\rm CO}$. Likewise, to understand the $\alphacii$–$Z$ anticorrelation, we must focus on PDRs and GMCs and hence on the spatially resolved $\alphaciir$.

\section{Spatially resolved $\alphaciir$ conversion factor}
\label{sec:spatiallyalpha}
 
By leveraging the detailed physical information at $\approx 30$ pc scales of \texttt{SERRA} galaxies, we now turn to a study on how local conditions influence the conversion factor. In Figure \ref{fig:maps} we show the map of $\alphaciir=$\Sg/\Scii~for Lantana, Adenia, and Malva.
By visually inspecting the maps, we note that the spatially resolved conversion factor shows a clear radial decline, decreasing from the periphery toward the center of the sources, where $Z$, $n$, and $\mathcal{M}$ are higher. The radial behavior of $\alphaciir$ resembles that derived in PHANGS \citep[see e.g.][]{sandstrom2013, leroy2025} for the CO-to-$\rm H_2$ at $z=0$, where the central high surface and volume density regions of local galaxies are characterized by low $\alpha_{\rm CO}$.

To separately quantify the impact of gas metallicity, density, and turbulence on the $\alphaciir$, in Figure \ref{fig:2d_pdfs} we plot the 2D mass-weighted and $\rm [CII]$-weighted Kernel Density Estimates (KDEs) of $\alphaciir$ as a function of the spatially resolved (pixel by pixel) $Z$, $n$, and Mach number $\mathcal{M}$ for Lantana, Adenia, and Malva. We find a clear anti-correlation between the conversion factor with each of these three quantities, albeit for Adenia we obtain a larger scatter than for the other two galaxies; this is due to the fact that Adenia is caught after a major merger \citep{rizzo2022}.
Overall, the highest redshift galaxy, Lantana, having the lowest metallicity and a plume of patches in its ISM with low ($\log \mathcal{M}\approx 0.3$) Mach number, has marginally higher $\alphacii$. We further note that the mass-weighted and $\rm [CII]$-weighted KDEs closely match each other, albeit the [CII]-weighted ones do not show up in the high-end of the $\alphaciir$ range. These large $\alphaciir$ values correspond to patches of the ISM that are intrinsically [CII] faint because they are characterized by low-density ($\log (n/\rm cm^{-3}) \leq 0.5$) and low metallicity ($\log (Z/Z_{\odot}) \leq -2$). These high $\alphaciir$ plumes account for $<15\%$ of the total gas mass in every galaxy.

\subsection{A physical intuition on how ISM properties shape $\alphaciir$}
\label{sec:physical_interpretation}

To gain a physical intuition on how the ISM properties shape $\alphaciir=\Sigma_{\rm gas}/\Sigma_{\rm [CII]}$ we exploit the [CII] emission model presented in \citet[][\citetalias{ferrara2019} hereafter]{ferrara2019} that isolates three key properties impacting $\Sigma_{\rm [CII]}$: metallicity, density, and the galaxy burstiness. \citetalias{ferrara2019} enables the computation of the surface brightness of [CII] from the ionized and the PDR layers of a gas slab illuminated by UV radiation from newly formed stars. The surface brightness is determined by the average gas density ($n$) of the HII/PDR environment -- characterized by electron density $n_e$ and neutral gas density $n_H$, respectively\footnote{Both $n_e$ and $n_H$ can be expressed as a function of $n$. In the ionized layer, $n_e=x_e n\approx n$ assuming an ionized fraction $x_e\approx 1$, while in the PDR the neutral gas density $n_H=(1-x_e) n \approx n$ given that $x_e\approx 0$.} --, the dust-to-gas ratio, ($\mathcal{D}\propto Z$, where $Z$ is the gas metallicity), and ionisation parameter, $U$. The latter, can be expressed in terms of observed quantities by deriving its relation ($U \propto$ \S*/\Sg, see eqs. 38 and 40 in \citetalias{ferrara2019}). This leaves us with the $\kappa_s$ parameter, describing the burstiness of the galaxy.

\begin{figure}
    \centering
    \includegraphics[scale=0.55]{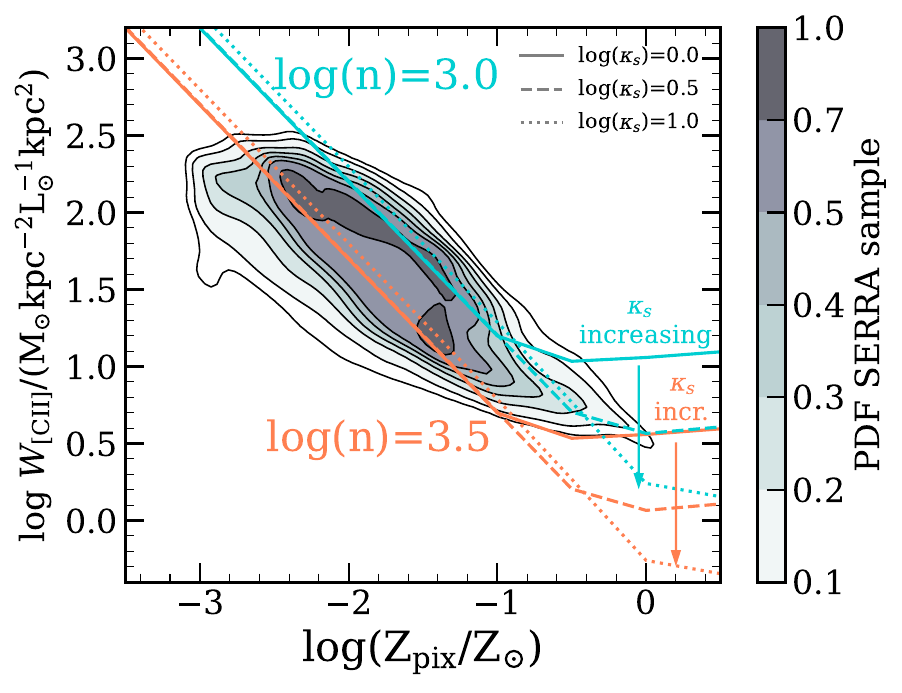}
    \caption{The 2D PDF of the spatially resolved conversion factor ($\alphaciir$) as a function of spatially resolved metallicity ($Z_{\rm pix}$), from the entire sample of \texttt{SERRA} sources. For comparison, we plot the theoretical estimates obtained with the \citetalias{ferrara2019} model assuming gas density $\log(n/\rm cm^{-3}) = 3$, in red, and $\log (n/\rm cm^{-3}) = 3.5$, in cyan. We plot results varying the burstiness parameter: $\log \kappa_s =0$ (solid lines), $\log \kappa_s=0.5$ (dashed lines), and $\log \kappa_s =1.0$ (dotted lines). }
    \label{fig:comparison_f19}
\end{figure}

In Figure \ref{fig:comparison_f19} we plot the $\alphaciir$ vs. $Z$ relation obtained by substituting the analytical equations for \Scii~by \citetalias{ferrara2019}. We vary also the gas density ($\rm \log(n/cm^{-3})=3,\, 3.5$, different colors), and the burstiness ($\log \kappa_s=0,0.5,1$, different lines). It is worth noting that the critical density of the [CII] line emission is $\log (n_{\rm crit}/\rm cm^{-3}) = 3.3$ \citep{tielens2005book}. In Fig. \ref{fig:comparison_f19} we note two distinct regimes: one at low metallicities, where $\alphaciir$ depends on $Z$ and $n$ but not on $\kappa_s$, and the other one at high metallicities where instead $\alphaciir$ is modulated by $\kappa_s$ and $n$, but is only weakly dependent on $Z$.
The first regime ($Z<0.2\,\rm Z_{\odot}$) can be understood by noticing that the [CII] flux can be written as (see \citetalias{ferrara2019} for details): 
 \begin{equation}
     \Sigma_{\rm [CII]} \propto n Z\Sigma_{\rm gas},
 \end{equation}
 hence, given that $\alphaciir = $\Sg/\Scii, it follows that $\alphaciir \propto (1/n Z)$ with no dependence on $\kappa_s$. At fixed $Z$, $\alphaciir$ decreases with increasing $n$ (see different colored lines in Fig. \ref{fig:comparison_f19}).

In the high-metallicity regime ($Z\gtrsim0.2 \rm Z_{\odot}$) instead, $\alphaciir$ reaches a constant value that depends on the burstiness. This can be understood by considering that, in this regime, $\Sigma_{\text{[CII]}}$ can be written as:
\begin{equation}
    \Sigma_{\text{[CII]}} \propto n Z \ln\left(1 + 10^5 w U \right)\,,
\end{equation}
with $w = (1 + 0.9 Z^{1/2})^{-1}$. Consider that i) $U$ can be linked to the \S* and $\kappa_s$ as
\begin{equation}
    U \propto 10^{-3} \kappa_s^{10/7} \Sigma_{\text{SFR}}^{-3/7}\,,
\end{equation}
and that ii) $\Sigma_{\text{gas}}$ can be expressed as a function of \S* and $\kappa_s$ via the KS relation
\begin{equation}
    \Sigma_{\text{gas}} = \left(\frac{\Sigma_{\text{SFR}}}{10^{-12}} \kappa_s\right)^{5/7}.
\end{equation}
Thus the final expression for $\alphaciir$ becomes:
\begin{equation}
    \alphaciir \propto \frac{\kappa_s^{-5/7}}{n Z \ln(1 + 10^2 w \kappa_s^{10/7})}\,,
\end{equation}
assuming \S* constant. At fixed $Z$ and $n$, for small values of $\kappa_s$ ($\kappa_s \approx 1$), the logarithm term goes as $\ln(1 + 10^2 w \kappa_s^{10/7}) \approx 10^2 w \kappa_s^{10/7}$, leading to:
\begin{equation}
    \alphaciir \sim  \kappa_s^{-15/7}\,.
\end{equation}
Thus, for small $\kappa_s$, $\alphaciir$ decreases as $\kappa_s^{-15/7}$.
For large $\kappa_s$ ($\kappa_s \approx 100$), the logarithm starts to be relevant, so the dominant scaling becomes:
\begin{equation}
    \alphaciir \sim \frac{\kappa_s^{-5/7}}{\ln \kappa_s}.
\end{equation}

This implies a slower decrease of $\alphaciir$ with increasing $\kappa_s$. Overall, bursty regions are expected to exhibit lower $\alphaciir$ values because the [CII] emission per unit mass is higher. This occurs as the molecular region behind the PDR shifts toward higher column densities, making more gas effectively capable of emitting [CII]. 
The dependence on metallicity $Z$ is still present as it is contained in both $w$ and the denominator of $\alphaciir$. It is possible to show that in this regime:
\begin{equation}
    \alphaciir \sim \frac{1}{Z \ln(1 + 10^2 w \kappa_s^{10/7})}.
\end{equation}
Thus, increasing $Z$ causes $\alphaciir$ to decrease
slower than a simple $1/Z$ dependence.
Obviously, the [CII] luminosity depends also on the gas density. At fixed metallicity, higher gas densities shift $\alphaciir$ to lower values, consistent with theoretical expectations (see Fig. \ref{fig:comparison_f19}).

Very interestingly, the \citetalias{ferrara2019} model accurately captures the average properties of the \texttt{SERRA} sample, with the 2D PDF of the simulated $\alphaciir$-$Z$ from all the pixels from all the galaxies in the sample nicely aligned with the theoretical lines. 
The turbulent and clumpy nature of the ISM is incorporated in \texttt{SERRA} by parameterizing the lognormal density distribution in each cell as a function of $\mathcal{M}$, which sets the dispersion $\sigma ^2 \approx {\rm ln}(1 + \mathcal{M}^2)$ of the lognormal. The average Mach number in our sample is $\langle \mathcal{M} \rangle = 15$ \citep[see also][]{pallottini2022}. The resulting $n_{\rm subgrid}$ is then used in the post-processing to compute the [CII] emission. Although a high-density power-law tail can develop due to self-gravity \citep{federrath2013, girichidis2014} and affect tracers of denser gas, its impact on [CII] is expected to be minimal, as the critical density of [CII] ($\approx 10^3\rm cm^{-3}$) is already well sampled by the lognormal distribution for $\mathcal{M}>10$ \citep[see][]{vallini2018}. For the entire \texttt{SERRA} sample, the [CII]-weighted $\langle \log (n_{\rm subgrid}/\rm cm^{-3}) \rangle =3.1$ is in excellent agreement with the $\log(n/\rm cm^{-3}) =[3.0,3.5]$ lines from the \citetalias{ferrara2019} model. The [CII]-weighted average gas density (from the simulation without subgrid implementation) is $\langle \log( n/\rm cm^{-3}) \rangle = 2.1$.

\subsection{Best fit equations for $\alphaciir$}

\begin{figure}
    \centering
    \includegraphics[scale=0.55]{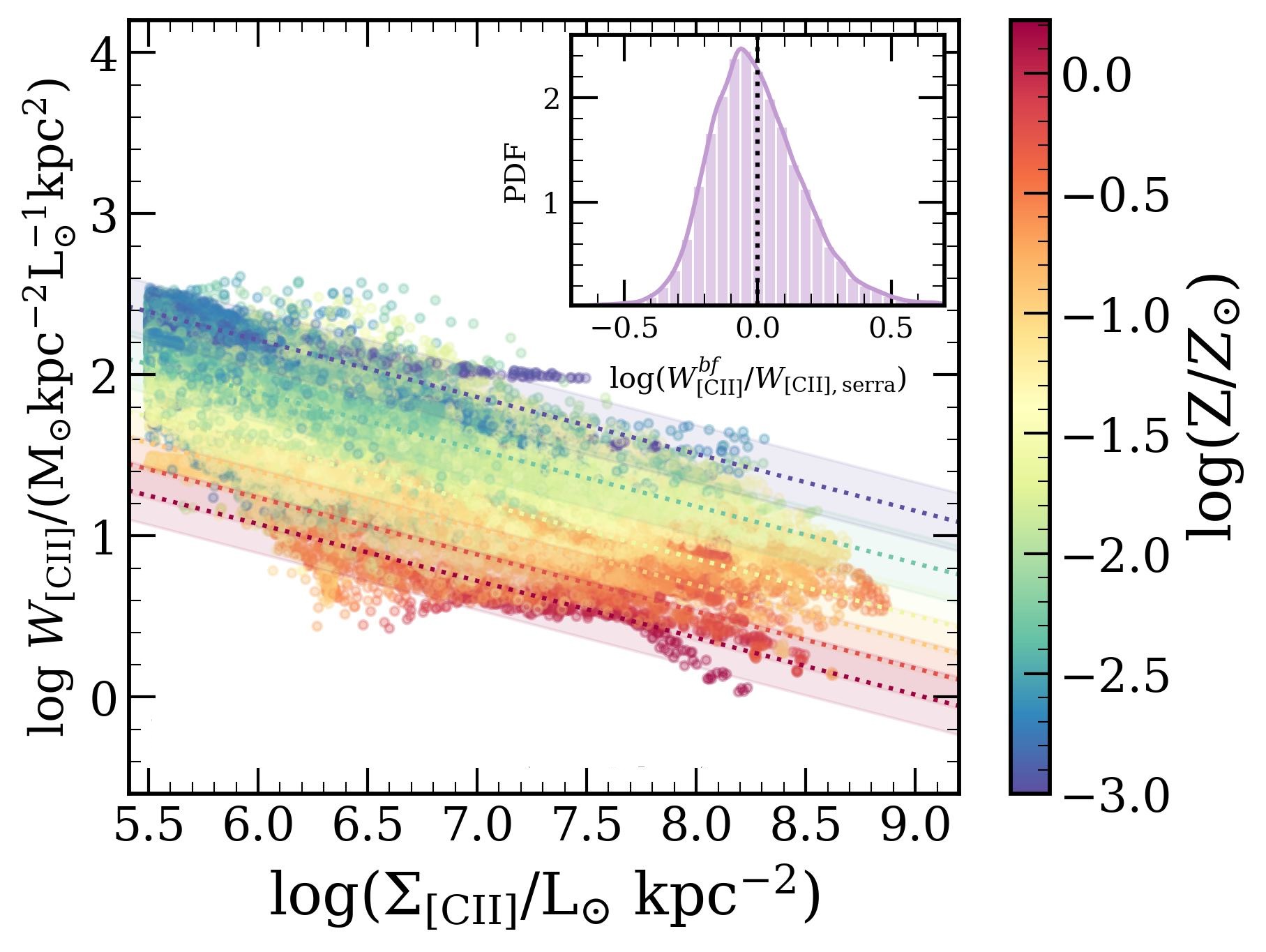}
    \caption{The spatially resolved conversion ($W_{\rm [CII]}$) factor as a function of the [CII] surface brightness \Scii~within $\approx30$ pc size pixels (colored points) in the ISM of \texttt{SERRA} galaxies. The color-code reflects the metallicity within each pixel. The best fit relation (dispersion) from Eq. \ref{eq:model}, for different metallicities, is plotted with dotted lines (shaded areas), using the same color-code adopted for \texttt{SERRA}. The inset shows the PDF of the log error between the $\alphaciir$ obtained pixel by pixel using the best fit relation, and the actual one from the simulation.}
    \label{fig:alphacii_sigmacii_bf}
\end{figure}

\begin{figure*}
    \centering
    \includegraphics[width=0.9\textwidth]{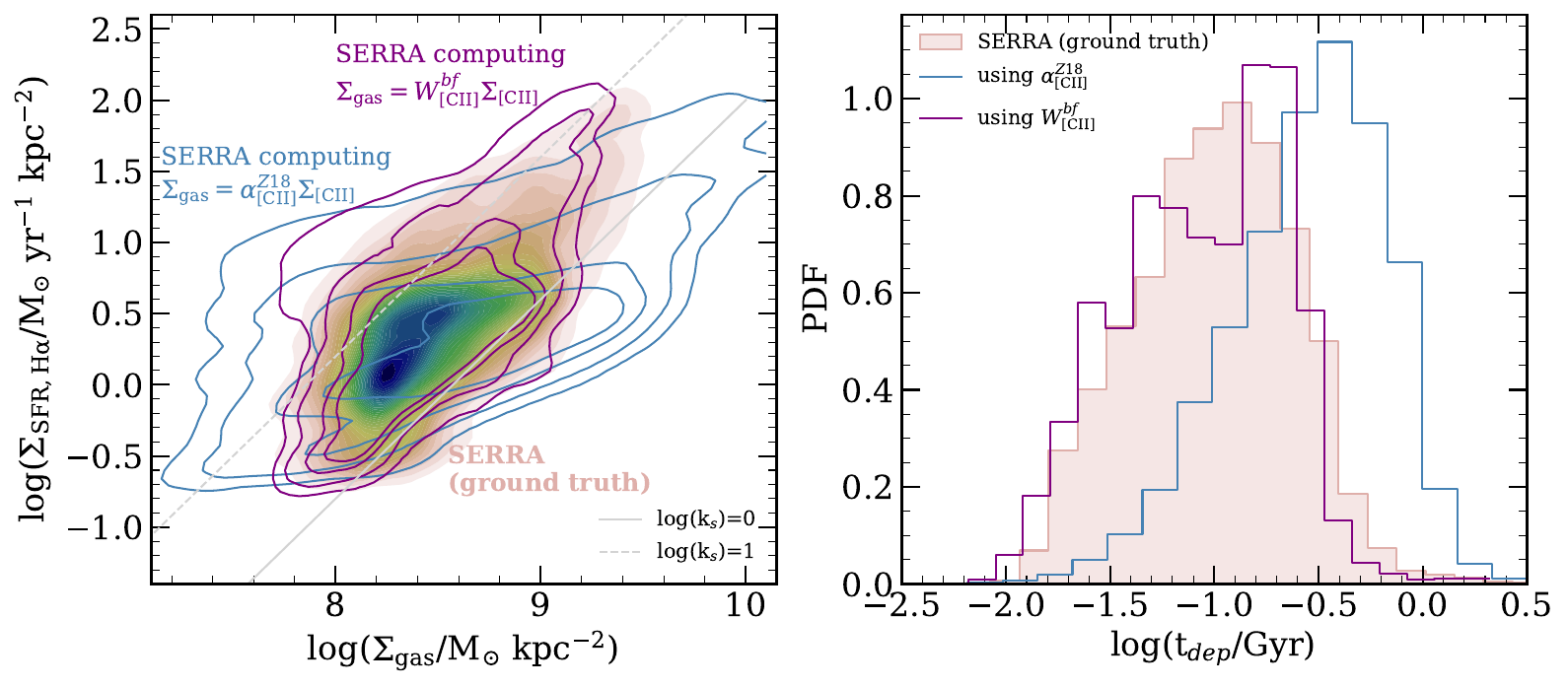}
    \caption{Left: The spatially resolved KS relation considering $\approx30\times$30 pc$^2$ pixels drawn from the entire sample of \texttt{SERRA} galaxies (color coded, ground truth). For each pixel the $\Sigma_{\rm SFR, H\alpha}$ is derived from the $H\alpha$ surface brightness using \citet{kennicutt1998}, and the $\Sigma_{\rm gas}$ is taken from the simulation. For comparison, we also report (blue contours) the inferred location of the pixels onto the KS plane assuming $\Sigma_{\rm gas}=\alphacii^{Z18}\Sigma_{\rm [CII]}$, namely adopting a constant \citet{zanella2018} conversion factor to get the gas surface density from $\Sigma_{\rm [CII]}$. The location in the KS plane resulting by using the best fit \emph{spatially resolved} conversion factor $W_{\rm [CII]}$ (eq. \ref{eq:model}) is plotted in purple. Right: The L$_{\rm [CII]}$-weighted distribution of the depletion time ($t_{\rm dep}$) in the \texttt{SERRA} pixels (shaded histogram), that inferred using $\alphacii ^{Z18}$ for getting the $\Sigma_{\rm gas}$ (blue), and that obtained using $W_{\rm  [CII]}$ from eq. \ref{eq:model}.}
    \label{fig:KS_relation_serra_pixels}
\end{figure*}

With this physical intuition at hand, we now turn to the \texttt{SERRA} sample and search for relations that are useful for inferring the conversion factor from spatially resolved observations of [CII]. For instance, an incorrect conversion factor might lead to a severe overestimate of \Sg~in the center of galaxies where the metallicity, density, and burstiness are overall higher than at the periphery. Stated differently,  $\alphaciir$ is expected to be overall lower than the global values.

To this aim we consider our \texttt{SERRA} sample, and we derive the relation between $W_{\rm  [CII]}$, measured on $\approx 30$ pc scales, and \Scii\, for all the galaxies in our sample ~(Fig.  \ref{fig:alphacii_sigmacii_bf}).
As expected, at fixed \Scii~the conversion factor increases for decreasing metallicity. Moreover, the brighter the \Scii~the lower is the conversion factor, indicating that brighter spots in the ISM of early galaxies correspond to denser PDRs, which are (see Figure \ref{fig:comparison_f19}) characterized by lower conversion factors with respect to regions with comparable metallicity but lower density.
Given that observationally the sub-pc density distribution in the ISM of high-$z$ galaxies is difficult to derive, we write a simplified, effective model that depends on \Scii~and $Z$
\begin{multline}
    \log (W_{\rm [CII]}/\rm M_{\odot} kpc^{-2}/ L_{\odot} kpc^{-2}) = \\ = A \log \left(\frac{\Sigma_{\rm [CII]}}{\rm L_{\odot}\, kpc^{-2}} \right) + B \log \left(\frac{Z}{Z_{\odot}}\right) + C + \sigma\,,
\label{eq:model}
\end{multline}
where $\sigma$ is a Gaussian noise term, quantifying the intrinsic scatter due to physical processes not explicitly captured by the metallicity nor the [CII] surface brightness.

We adopt a Bayesian framework and estimate the posterior distribution of the parameters $A$, $B$, $C$ and the intrinsic scatter $\sigma$, using a Markov chain Monte Carlo (MCMC) integrator \citep[\texttt{emcee};][and references therein]{foreman2013}\footnote{We run \texttt{emcee} with
10 random walkers exploring the parameter space for $50000$ chain steps}. We assume flat priors $A<0$, $B<0$, and $\sigma >0$. The resulting posteriors are $A= -0.355^{0.003}_{-0.017}$, $B= -0.324^{0.048}_{-0.004}$, $C = 3.37^{0.191}_{-0.026}$, and $\sigma =0.184$. These are summarized in Table \ref{tab:best_fit_relations}. Substituting the best-fit parameters in eq. \ref{eq:model}, the resulting relations (scatter) at fixed metallicity are plotted with lines and shaded areas in Fig. \ref{fig:alphacii_sigmacii_bf}. We analyze the probability density function (PDF) of the logarithmic error, defined as the difference between $\log \alphaciir$ from the simulation and the value predicted by the best-fit equation (Eq. \ref{eq:model}). We find that this distribution mostly falls within $\pm$0.5 dex, indicating that the model's predictions generally agree with the simulation within this margin of error.
Finally, if the information of the spatially resolved metallicity is not available, in  Table \ref{tab:best_fit_relations}, we provide the parameters of the linear regression: 
\begin{equation}
  \log (W_{\rm [CII]}/\rm M_{\odot} kpc^{-2}/ L_{\odot} kpc^{-2}) = A \log  \left(\frac{\Sigma_{\rm [CII]}}{\rm L_{\odot}\, kpc^{-2}} \right) + C  \,.
 \label{eq:wcii_sigma}
\end{equation}
The important caveat is that the scatter around this relation, $\sigma=0.4$, is wider than that obtained for the physical model (eq. \ref{eq:model}), since the metallicity $Z$ is not explicitly accounted for.

\section{Discussion}
\label{sec:discussion}

\subsection{Using $\alphaciir$ to derive the KS relation and $t_{\rm dep}$}
\begin{figure*}
    \centering
    \includegraphics[scale=0.5]{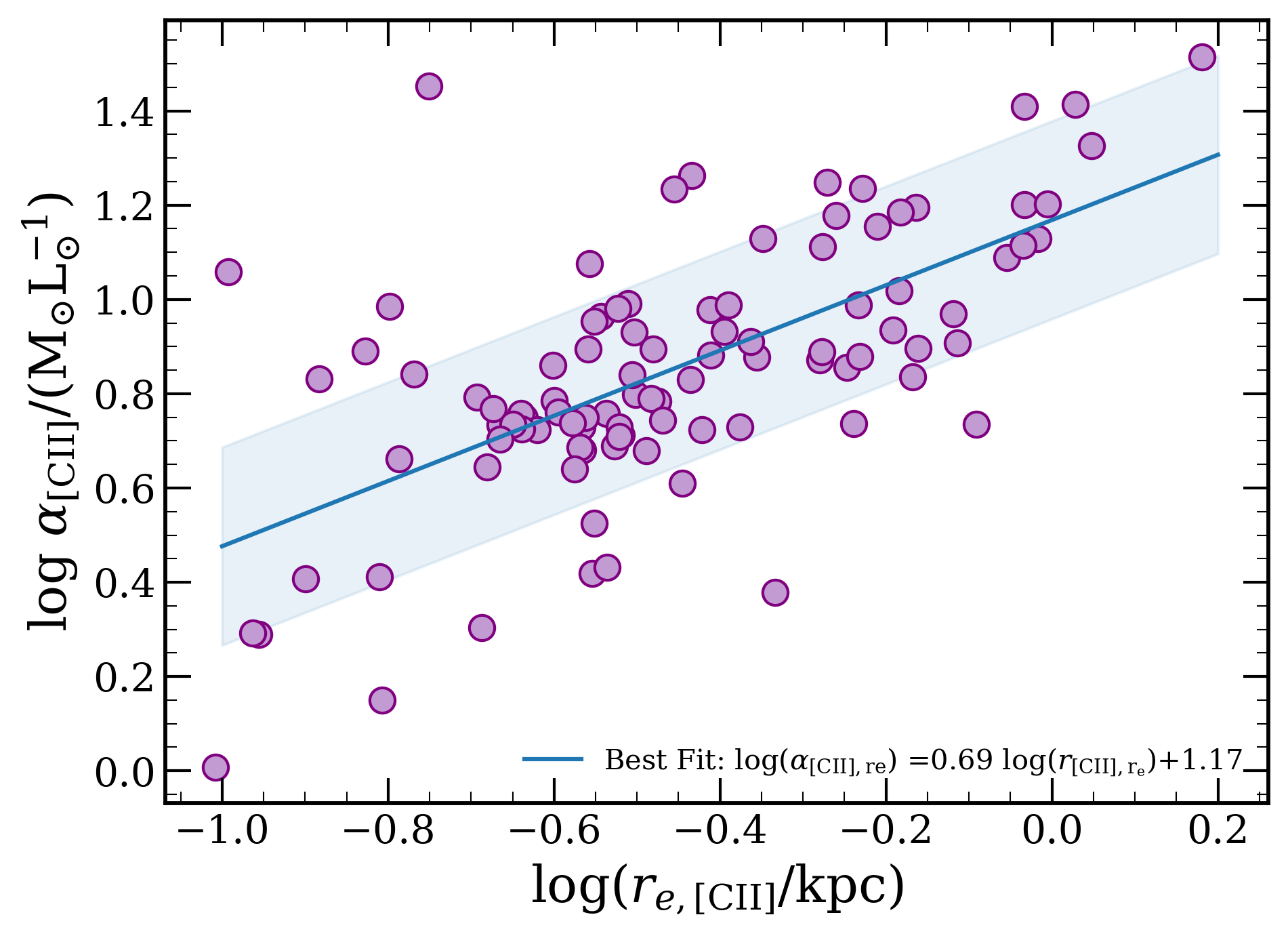}
    \includegraphics[scale=0.5]{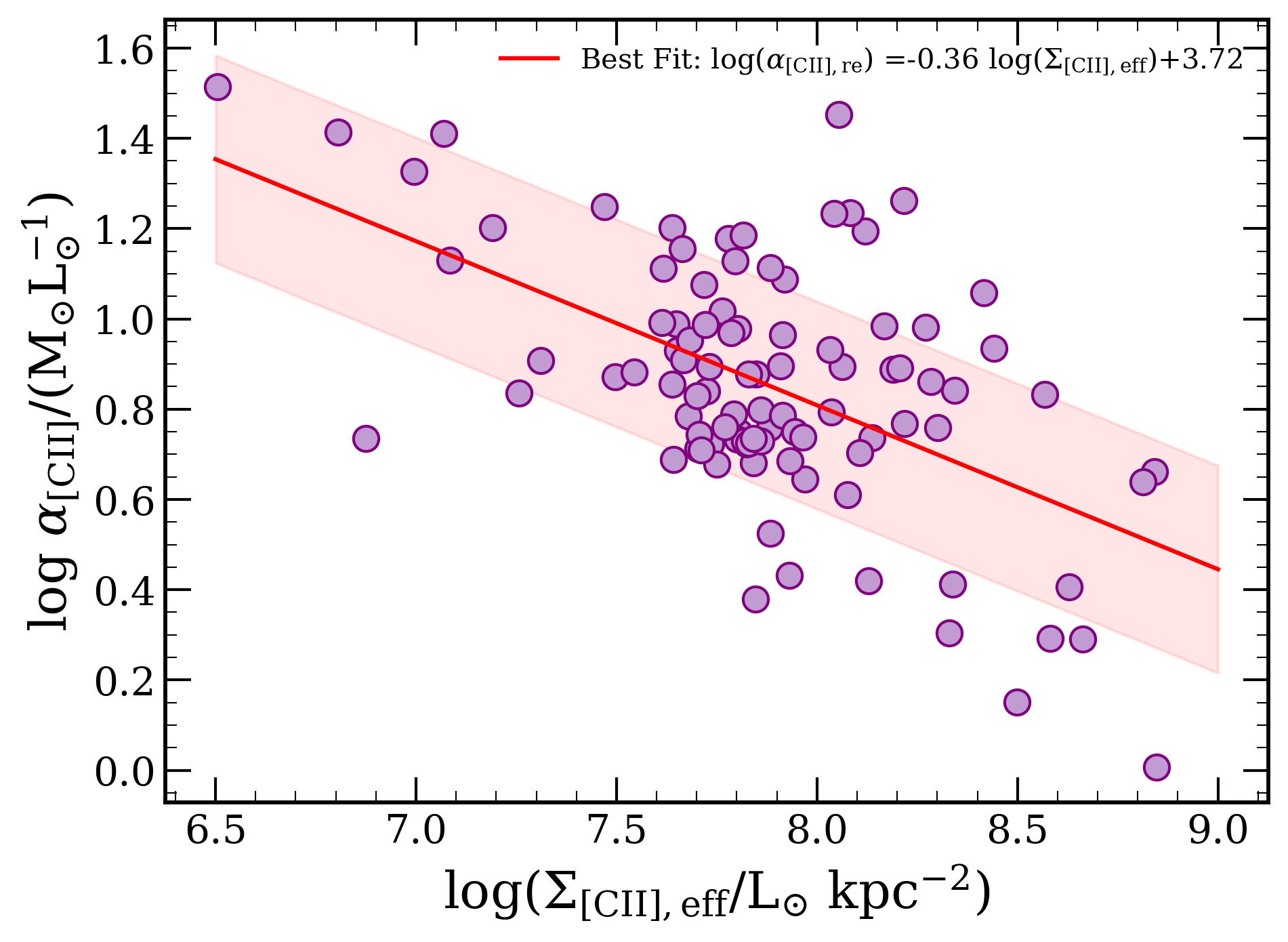}
    \caption{Left: The $\alphacii$ as a function of the [CII] half light radius for the sample considered in this work. Right: The $\alphacii$ as a function of the [CII] effective surface brightness.}
    \label{fig:thresholdeffect}
\end{figure*}
In the previous Sections, we showed that the spatially resolved $W_{\rm [CII]}$ exhibits a clear decreasing trend toward the central regions of galaxies, which correspond to the most actively star-forming, metal-enriched, and densest patches of their ISM. These regions are also characterized by the highest [CII] surface brightness, with which $\alphaciir$ anticorrelates (see Eq. \ref{eq:model}). In what follows, we quantify the bias introduced by using a fixed $\alpha_{\rm [CII]}$ instead of $\alphaciir$ when determining the slope and normalization of the KS relation and the gas depletion time.

To do so, in Figure \ref{fig:KS_relation_serra_pixels}, we start with the spatially resolved KS relation obtained from \texttt{SERRA}, considering 30$\times$30 pc$^2$ pixels drawn from the entire sample. For each pixel, $\Sigma_{\rm SFR, H\alpha}$ is derived from the H$\alpha$ surface brightness using \citet{kennicutt1998}, while $\Sigma_{\rm gas}$ is taken directly from the simulation. As shown by the color-coded shaded area in Figure \ref{fig:KS_relation_serra_pixels}, the 2D PDF of all \texttt{SERRA} pixels aligns well with the KS relation, consistent with the standard $\approx 1.4$ slope, with pixels characterized by burstiness parameters in the range $\kappa_s=1-10$.
However, if instead of using $\Sigma_{\rm gas}$ from the simulation, we adopt a fixed $\alpha_{\rm [CII]}$ \citep{zanella2018} to convert $\Sigma_{\rm [CII]}$ into $\Sigma_{\rm gas} = \alphacii \Sigma_{\rm [CII]}$, the resulting KS relation becomes significantly flatter. This occurs because the gas surface density is overestimated in regions with high $\Sigma_{\rm [CII]}$, highlighting a critical caveat when using spatially resolved [CII] observations to study the KS relation (see also Accard et. al, in prep. for the ALPINE-CRISTAL sample). In contrast, when we apply our best-fit $W_{\rm [CII]}$–$\Sigma_{\rm [CII]}$ relation (Eq. \ref{eq:model}) to convert $\Sigma_{\rm [CII]}$ into $\Sigma_{\rm gas}$, we successfully recover the intrinsic KS relation from \texttt{SERRA}, reinforcing the validity of Eq. \ref{eq:model} and its use in spatially resolved observations, both at high-$z$ and in the local Universe \citep[see the recent][]{kovavic2025}.

Furthermore, adopting a single conversion factor leads to an overestimation of the depletion time ($t_{\rm dep}$) in bursty ISM regions and, consequently, across entire galaxies. This is depicted in the right panel of Fig. \ref{fig:KS_relation_serra_pixels} where we plot the [CII] weighted PDF of the depletion times inferred pixel by pixels inverting the KS relation obtained with the three different approaches outlined before. While in the SERRA galaxies (labelled as ground truth in Fig. \ref{fig:KS_relation_serra_pixels}) the average depletion time is $\langle t_{\rm dep} \rangle \approx 0.1 \rm \, Gyr$, when using the fixed conversion factor one would infer a 4$\times$ higher depletion time, $\langle t_{\rm dep} \rangle \approx 0.4 \rm \, Gyr$.

\subsection{The impact of compactness on the global $\alpha_{\rm [CII]}$}

As discussed in Sec.~\ref{sec:physical_interpretation}, $W_{\rm [CII]}$ anticorrelates with $n$, $Z$, and burstiness parameter $\kappa_s$. Given that these properties also drive an increase in $\Sigma_{\rm [CII]}$, regions with higher surface brightness tend to have lower $W_{\rm [CII]}$. This trend, observed at spatially resolved scales, can also influence the global $\alpha_{\rm [CII]}$.  

If a galaxy as a whole is characterized by a higher effective [CII] surface brightness, $\Sigma_{\rm [CII], eff}= L_{\rm [CII]}/(2\pi r_{e,\rm [CII]}^2)$, this implies that a large fraction of its ISM consists of bright patches with low $W_{\rm [CII]}$. Since $\alpha_{\rm [CII]}$ can be considered a [CII]-luminosity weighted mean of $W_{\rm [CII]}$, this suggests that more compact galaxies with higher $\Sigma_{\rm [CII], eff}$ are also likely to have lower global $\alpha_{\rm [CII]}$. Shedding light on this connection is key for interpreting unresolved observations and for accurately estimating gas masses from $L_{\rm [CII]}$.  
From the 5$\times$5 kpc$^2$ maps we computed half-light radius ($r_{e,\rm [CII]}$) of the [CII] emission to derive $\Sigma_{\rm [CII], eff}$. We then calculated the conversion factor within $r_{e,\rm [CII]}$ as $\alpha_{\rm [CII]} = M_{\rm gas, r_e} / L_{\rm [CII], r_e}$.
In Figure \ref{fig:thresholdeffect}, we present $\alpha_{\rm [CII]}$ as a function of $r_{e,\rm [CII]}$, and $\Sigma_{\rm [CII], eff}$.
We find, as expected, that galaxies with higher [CII] surface brightness—or equivalently, more compact galaxies with smaller $r_{e,\rm [CII]}$—tend to exhibit lower values of $\alpha_{\rm [CII]}$. This result highlights the impact of galaxy structure on the global conversion factor and suggests that $\Sigma_{\rm [CII], eff}$, reminiscent of the local \Scii, is a good indicator of the conversion factor that should be used for a more precise estimate.
We fit the following logarithmic relations:
\begin{equation}
\log (\alpha_{\rm [CII]}/\rm M_{\odot} L_{\odot}^{-1}) = A\log (r_{e,\rm [CII]}/kpc)  + C,
\label{eq:alphare}
\end{equation}
and:
\begin{equation}
\log (\alpha_{\rm [CII]}/\rm M_{\odot} L_{\odot}^{-1}) = A \log (\Sigma_{\rm [CII], eff}/\rm L_{\odot} kpc^{-2}) + C.
\label{eq:alphasigma}
\end{equation}
The parameters, along with their errors are listed in Table \ref{tab:best_fit_relations}.

\section{Conclusions}
\label{sec:conclusions}

In this paper, we studied the [CII]-to-gas conversion factor in high-redshift ($z\approx 4-9$) galaxies extracted from the \texttt{SERRA} cosmological zoom-in simulation, both globally and down to spatially resolved scales. Our main findings can be summarized as follows:

\begin{itemize}
   \item[$\bullet$] The mean global conversion factor in the \texttt{SERRA} sample is $\log(\alphacii/\rm M_{\odot}\,L_{\odot}^{-1}) = 1.28$ ($\sigma = \pm 0.2 \rm$ dex) within the ranges derived in previous works, both observationally and with simulations. The $\alphacii$ anticorrelates with $Z$, and we provide a best fit relation in Table. \ref{tab:best_fit_relations}, while no significant correlation is found with redshift, stellar mass, or deviation from the MS.\\

    \item[$\bullet$] The spatially resolved conversion factor, $W_{\rm [CII]}$, anticorrelates $n$, and $Z$. For $Z>0.2Z_{\odot}$, also the galaxy burstiness modulates the $W_{\rm [CII]}$, with bursty patches being characterized by lower conversion factors. Since these regions are also characterized by high [CII] surface brightness, this leads to an inverse correlation between $W_{\rm [CII]}$ and $\Sigma_{\rm [CII]}$. In Tab. \ref{tab:best_fit_relations} we report the best-fit $W_{\rm [CII]}$ as a function of $Z$ and $\Sigma_{\rm [CII]}$, which is a tight ($1\sigma =0.18$ dex) relation, and the $W_{\rm [CII]}$ vs $\Sigma_{\rm [CII]}$ only, which instead has a large dispersion ($1\sigma=0.4$ dex). Either of the two can be used to derive the cold gas surface density from spatially resolved [CII] maps, depending on whether information on $Z$ is available from other tracers.\\
    
    \item[$\bullet$] Adopting a fixed global $\alpha_{\rm [CII]}$ to infer \Sg~in the spatially resolved Kennicutt-Schmidt (KS) relation can lead to an artificially flat slope and to an overestimation of the depletion time up to $\approx 4 \times$. This is because applying a uniform conversion factor overestimates $\Sigma_{\rm gas}$ in bright $\Sigma_{\rm [CII]}$ regions, where the actual conversion factor is lower.\\
    
    \item[$\bullet$] The global $\alpha_{\rm [CII]}$ is influenced by galaxy compactness. More compact galaxies (i.e., those with smaller $r_{e, \rm [CII]}$ and higher $\Sigma_{\rm [CII], eff}$) tend to be characterized by lower global $\alpha_{\rm [CII]}$. Our work hence suggests that the non detection of [CII] in compact JWST selected galaxies at $z>10$ puts very strong constraints on the cold gas mass, pointing towards very short depletion times\\
\end{itemize}   
With this work, we provide ready to use physically motivated relations (gathered in Table \ref{tab:best_fit_relations}) that allow us to account for spatial variations in $\alpha_{\rm [CII]}$. This is crucial for estimating gas masses and surface densities in both resolved and unresolved observations with ALMA in the EoR. We stress that using a physically motivated conversion factor can help in pinning down the right observation time required for following up in [CII] JWST selected galaxies which are often metal poor and relatively compact.
\begin{table*}
    \caption{Best-fit relations for spatially resolved and global [CII] conversion factors. The fits follow the form $Y = A X_1 + B X_2 + C$ (for two-variable fits) or $Y = A X_1 + C$ (for single-variable fits). The units are as follows: $\alphacii$ in [$\rm M_{\odot} L_{\odot}^{-1}$], $\alphaciir$ in [$(\rm M_{\odot} kpc^{-2})/(L_{\odot} kpc^{-2})$], $Z$ in [$Z_{\odot}$], $\Sigma_{\rm [CII]}$ in [$\rm L_{\odot} kpc^{-2}$], and $\rm r_{e, \rm [CII]}$ in [kpc]. We include the 1$\sigma$ dispersion of residuals around every best-fit relation. The last column lists the reference equation in the main text of the paper.}
    \centering
    \label{tab:best_fit_relations}
    \renewcommand{\arraystretch}{1.2}
    \setlength{\tabcolsep}{10pt}    
    \begin{tabular}{l l l c c c c c}
        \toprule
        $Y$ (best fit) & $X_1$ & $X_2$ (if present) & \textbf{$A$} & \textbf{$B$} & $\textbf{$C$}$ & \textbf{$\sigma$} & Eq. \\
        \midrule
        \multicolumn{8}{c}{\textbf{Global Relations}} \\
        \midrule
        $\log \alpha_{\rm [CII]}$ & $\log Z$ & — & $-0.39 \pm 0.04$ & — & $0.67 \pm 0.06$ & $0.14$ & [\ref{eq:alpha_Z_fit}] \\
        $\log \alpha_{\rm [CII]}$ & $\log r_{e, \rm [CII]}$ & — & $0.69 \pm 0.08$ & — & $1.17 \pm 0.04$ & $0.21$ & [\ref{eq:alphare}] \\
        $\log \alpha_{\rm [CII]}$ & $\log \Sigma_{\rm [CII], eff}$ & — & $-0.36 \pm 0.06$ & — & $3.7 \pm 0.5$ & $0.23$ & [\ref{eq:alphasigma}]\\
        \midrule
        \multicolumn{8}{c}{\textbf{Spatially Resolved Relations}} \\
        \midrule
        $\log W_{\rm [CII]}$ & $\log \Sigma_{\rm [CII]}$ & $\log Z$  & $-0.355^{+0.003}_{-0.017}$& $-0.324^{+0.048}_{-0.004}$ & $3.37^{+0.191}_{-0.026}$ & $0.18$ & [\ref{eq:model}] \\
        $\log W_{\rm [CII]}$ & $\log \Sigma_{\rm [CII]}$ & — & $-0.51 \pm 0.001$ & — & $4.93 \pm 0.008$ & $0.40$ & [\ref{eq:wcii_sigma}]\\
        \bottomrule
    \end{tabular}
\end{table*}

\begin{acknowledgements}
We thank the anonymous reviewer for their constructive comments that improved the quality of this work. LV acknowledges support from the INAF Minigrant "RISE: Resolving the ISM and Star formation in the Epoch of Reionization" (Ob. Fu. 1.05.24.07.01).
We gratefully acknowledge the support of the Lorentz Center for the organization of the workshop "Synergistic ALMA+JWST view of the early Universe", held in December, 2024 in Leiden, during which this work was conceived.
We acknowledge the CINECA award under the ISCRA initiative for the availability
of high-performance computing resources and support from the Class B
project SERRA HP10BPUZ8F (PI: Pallottini).
We gratefully acknowledge the computational resources of the Center for High-Performance Computing
(CHPC) at Scuola Normale Superiore.
AZ acknowledges support from the INAF Minigrant "Clumps at cosmological distance: revealing their formation, nature, and evolution" (Ob. Fu. 1.05.23.04.01) and from the European Union – NextGeneration EU within PRIN 2022 project n.20229YBSAN - "Globular clusters in cosmological simulations and in lensed fields: from their birth to the present epoch".
CG, FP, and LV acknowledge funding from the ASI-INAF contract n. 2023-15-Q.0 CUP F83C25000230001 “Attività scientifiche propedeutiche alla partecipazione alla missione PRIMA del JPL/NASA".
We acknowledge use of Astropy \citep{astropy2018}, Matplotlib \citep{matplotlib2007}, SciPy \citep{scipy2020}, Seaborn \citep{seaborn2020}, Pandas \citep{pandas2020}. 
\end{acknowledgements}

%
\bibliographystyle{aa_url_style}  
\bibliography{bibliography}
\appendix
\section{Global trends in SERRA}\label{sec:appendix}
\begin{figure}[h]
\centering
    \includegraphics[width=0.45\textwidth]{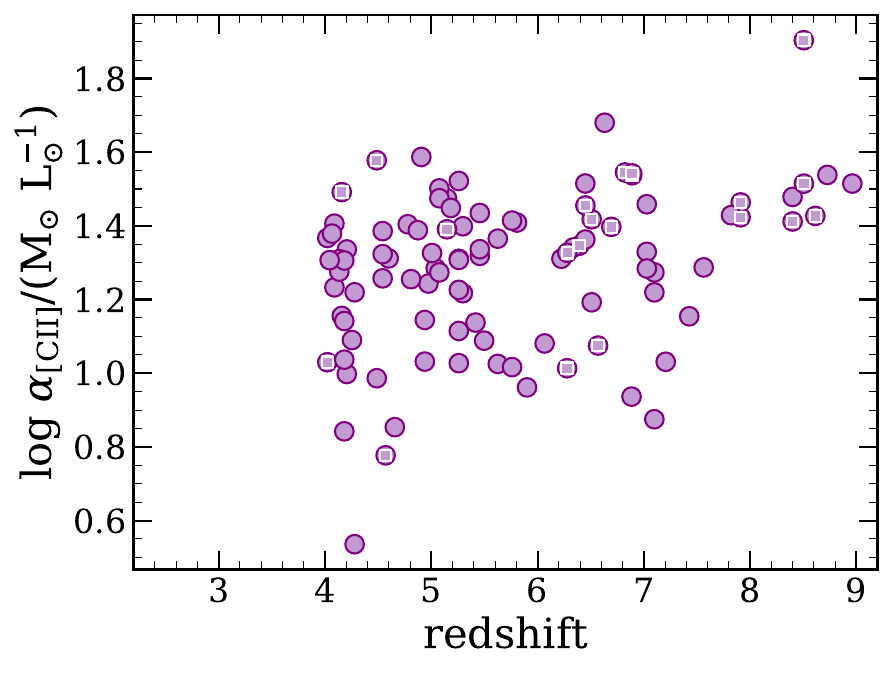}
    \includegraphics[width=0.45\textwidth]{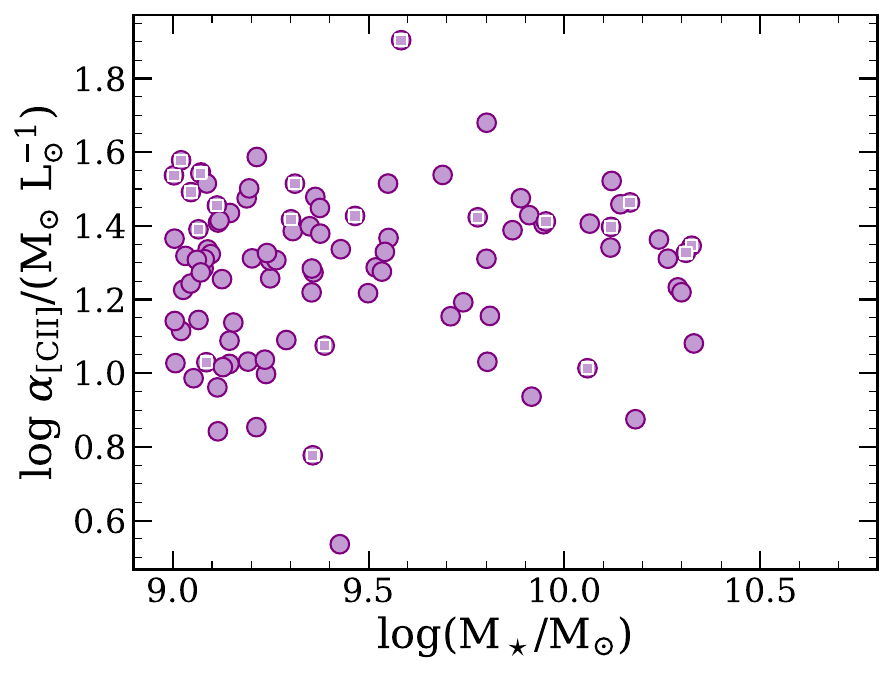}
    \includegraphics[width=0.45\textwidth]{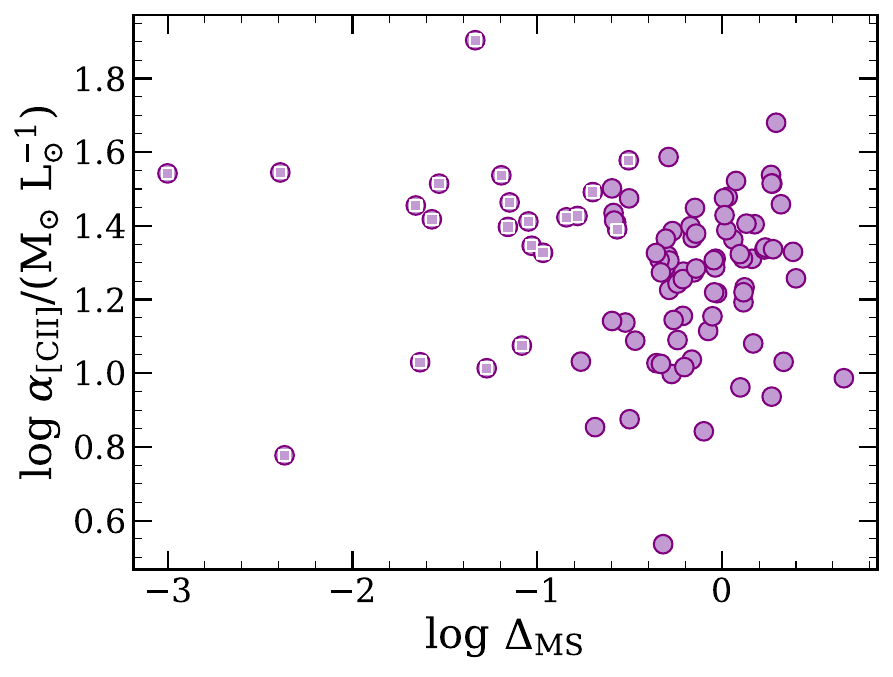}
    \caption{The global $\alphacii$ vs redshift (top panel), stellar mass (middle panel), and deviation from the MS (bottom panel) for the \texttt{SERRA} sample. We adopt the MS by \citet{cole2025} in the corresponding redshift bin of each SERRA galaxy. The meaning of symbols is the same outlined in Fig. \ref{fig:sample}.}
    \label{fig:nocorrelation}
\end{figure}

In Fig. \ref{fig:nocorrelation} we report the scatter plot between $\alphacii$ and redshift, stellar mass, and deviation with respect to the main sequence.
%
\end{document}